\documentclass[preprint,12pt,a4paper]{elsarticle}

\makeatletter
\def\ps@pprintTitle{%
  \let\@oddhead\@empty
  \let\@evenhead\@empty
  \def\@oddfoot{\reset@font\hfil\thepage\hfil}
  \let\@evenfoot\@oddfoot
}
\makeatother

\usepackage{amssymb}

\usepackage{hyperref}
\usepackage{breakurl}

\usepackage[fleqn]{amsmath}
\begin{document}

\begin{frontmatter}




\title{Extrapolation of Turbulence Intensity Scaling to $Re_{\tau} \gg 10^5$}


\author[au1]{Nils T. Basse}
\ead{nils.basse@npb.dk}

\address[au1]{Independent Scientist \\ Trubadurens v\"ag 8, 423 41 Torslanda, Sweden \\ \vspace{10 mm} \small {\rm \today}}

%
%
%
%
%
%

\begin{abstract}
We have characterized a transition of turbulence intensity (TI) scaling for friction Reynolds numbers $Re_{\tau} \sim 10^4$ in the companion papers [Basse, N.T. Scaling of global properties of fluctuating and mean streamwise velocities in pipe flow: Characterization of a high Reynolds number transition region. {\it Phys. Fluids} {\bf 2021}, {\it 33}, {\it 065127}] and [Basse, N.T. Scaling of global properties of fluctuating streamwise velocities in pipe flow: Impact of the viscous term. {\it Phys. Fluids} {\bf 2021}, {\it 33}, {\it 125109}]. Here, we build on those results to extrapolate TI scaling for $Re_{\tau} \gg 10^5$, under the assumption that no further transitions exist. Scaling of the core, area-averaged and global peak TI demonstrates that they all scale inversely with the logarithm of $Re_{\tau}$, but with different multipliers. Finally, we confirm the prediction that the TI squared is proportional to the friction factor for $Re_{\tau} \gg 10^5$.
\end{abstract}

\end{frontmatter}



\section{Introduction}

Turbulence intensity (TI) is important, both as a boundary condition (BC) for computational fluid dynamics (CFD) simulations \cite{greenshields_a} and for industrial applications such as quantifying repeatability of flowmeter measurements \cite{baker_a}.

A transition of streamwise velocity in pipe flow for friction Reynolds numbers $Re_{\tau}=\delta U_{\tau}/\nu \sim 10^4$ exists and has been characterized, both for the mean and fluctuating component \cite{basse_a,basse_b}. Here, $\delta$ is the boundary layer thickness (pipe radius $R$ for pipe flow), $U_{\tau}$ is the friction velocity and $\nu$ is the kinematic viscosity. As was also demonstrated, the transition implies a corresponding transition for the TI scaling. We speculate that the transition may be understood as a parallel to the "drag crisis", i.e. a sudden drop in the drag coefficient above a certain Reynolds number \cite{prandtl_a}.

The Princeton Superpipe measurements \cite{smits_a,hultmark_a} have an upper limit of $Re_{\tau} \sim 10^5$; in this paper, our goal is to use results based on these measurements to extrapolate scaling behavior for flows where $Re_{\tau} \gg 10^5$. For this exercise, we assume that no further transitions take place in addition to the one we have discussed.

Flows with higher Reynolds numbers than those in the Princeton Superpipe have been measured in atmospheric surface layers (ASL), $Re_{\tau} \sim 6 \times 10^5$ \cite{hutchins_a}, and experiments with superfluid Helium II have reached bulk Reynolds numbers $Re_D=D \langle U_{\rm g,mean} \rangle_{\rm AA}/\nu \sim 10^7-10^8$ \cite{fuzier_a,saint-michel_a,mastracci_a}, similar to the ASL range. Here, $D=2R$ is the pipe diameter and $\langle U_{\rm g,mean} \rangle_{\rm AA}$ is the area-averaged (AA) mean velocity, where the subscript "g" indicates "global". An industrial application where high Reynolds numbers may occur is cooling of superconducting magnets with superfluid Helium II \cite{lebrun_a}. For even higher present-day Reynolds numbers an example is accretion disks, where $Re_D$ can reach values as high as $10^{15}$ \cite{curtis_a}. In the early universe, bulk Reynolds numbers of up to $10^{16}$ have been estimated \cite{giovannini_a}. Thus, in addition to extreme industrial applications, our findings for extrapolated TI scaling are applicable to flow phenomena such as those observed in meteorology and cosmology.

Our main goal is to quantify the behavior of extrapolated TI scaling, mainly (i) scaling with $Re_{\tau}$, (ii) differences between the core, AA and global peak TI and (iii) the relationship between TI and the friction factor $\lambda=8 \times U_{\tau}^2/\langle U_{\rm g,mean}^2 \rangle_{\rm AA}$.

The paper is organized as follows: In Section \ref{sec:asymp}, asymptotic scaling expressions are derived, both for fluctuating and mean velocities. The resulting TI scalings are presented in Section \ref{sec:ti} and an associated discussion has been placed in Section \ref{sec:disc}. Finally, we conclude on our findings in Section \ref{sec:conc}.

\section{Asymptotic Scaling expressions}
\label{sec:asymp}

\subsection{Velocity Fluctuations}

We use an equation for the square of the normalised fluctuating velocity $u$ including the viscous term $V$ as formulated in \cite{perry_a}:

\begin{eqnarray}
\label{eq:u_rms_sq}
\label{eq:fluc_sq_perry}
 \frac{{\overline{u^2_{\rm g,fluc}}}(z)}{U_{\tau}^2} &=& B_{\rm g,fluc} - A_{\rm g,fluc} \log (z/\delta) - C_{\rm g,fluc} (z^+)^{-1/2} \\
   &=& B_{\rm g,fluc} - A_{\rm g,fluc} \log (z/\delta) + V(z^+),
\end{eqnarray}

\noindent where the subscript "fluc" indicates "fluctuating". Overbar is time averaging, $z$ is the distance from the wall and $z^+=z U_{\tau}/\nu$ is the normalized distance from the wall. Note that $z/\delta = z^+/Re_{\tau}$.

Asymptotic values for the fit parameters as derived in \cite{basse_b} are used:

\begin{eqnarray}
  \lim_{Re_{\tau}\to\infty} A_{\rm g,fluc} &=& 1.60 \\
  \lim_{Re_{\tau}\to\infty} B_{\rm g,fluc}  &=& 0.96 \\
  \lim_{Re_{\tau}\to\infty} C_{\rm g,fluc}/\sqrt{Re_{\tau}} &=& 0.12 \label{eq:C_g_div}
\end{eqnarray}

In the remainder of the paper we use the asymptotic fit parameter values so we will leave the prefix "$\lim_{Re_{\tau}\to\infty}$" out.

The location of the peak value of the fluctuating velocity is:

\begin{equation}
\frac{z}{\delta} \bigg \rvert_{\rm peak} = \left[ \frac{C_{\rm g,fluc}}{2 A_{\rm g,fluc} \sqrt{Re_{\tau}}} \right]^2 = 1.41 \times 10^{-3},
\end{equation}

\noindent see also Equation (30) in \cite{basse_b}. This can be reformulated to state that the peak is located 0.141 \% of the pipe radius from the wall, i.e. 141 $\mu$m from the wall for a pipe with a 100 mm radius. The peak was called the "global peak" in \cite{basse_b}, and it is a combination of the inner and outer peaks of the velocity fluctuations \cite{smits_b}. However, for very high Reynolds numbers, the peak is dominated by the outer peak.

The magnitude of the fluctuating velocity in the core, i.e. for $z/\delta=1$, is:

\begin{equation}
\frac{{\overline{u^2_{\rm g,fluc}}}(z/\delta=1)}{U_{\tau}^2} = B_{\rm g,fluc}-\frac{C_{\rm g,fluc}}{\sqrt{Re_{\tau}}} = 0.84
\end{equation}

The corresponding magnitude of the peak is given by:

\begin{equation}
\frac{{\overline{u^2_{\rm g,fluc}}}(z/\delta \rvert_{\rm peak})}{U_{\tau}^2} = B_{\rm g,fluc}-2A_{\rm g,fluc} \log \left( \frac{C_{\rm g,fluc}}{2 A_{\rm g,fluc} \sqrt{Re_{\tau}}} \right) - 2 A_{\rm g,fluc} = 8.27,
\end{equation}

\noindent a slight correction to the value 8.20 presented as Equation (35) in \cite{basse_b}.

The area-averaged (AA) value of the fluctuating velocity is:

\begin{equation}
\biggl \langle \frac{{\overline{u^2_{\rm g,fluc}}}}{U_{\tau}^2} \biggr \rangle_{\rm AA} = B_{\rm g,fluc} + \frac{3}{2} A_{\rm g,fluc} - \frac{8 C_{\rm g,fluc}}{3 \sqrt{Re_{\tau}}} = 3.04
\end{equation}

We conclude that the fluctuating velocity does not scale with $Re_{\tau}$.

\subsection{Mean Velocity}

The normalized mean velocity $U$ is given by \cite{basse_a}:

\begin{equation}
\frac{U_{\rm g,mean}(z)}{U_{\tau}} = \frac{1}{\kappa_{\rm g,mean}} \log(z^+) + A_{\rm g,mean},
\end{equation}

\noindent where $A_{\rm g,mean}=1.01$ and $\lim_{Re_{\tau}\to\infty} \kappa_{\rm g,mean} = 0.34$; note that this value is quite close to the value of 1/3 found in \cite{tennekes_a}.

The core value of the normalized mean velocity squared is:

\begin{equation}
\label{eq:mean_core}
\lim_{Re_{\tau}\to\infty} \frac{U_{\rm g,mean}^2(z/\delta=1)}{U_{\tau}^2} = \frac{\log^2(Re_{\tau})}{\kappa_{\rm g,mean}^2} = 8.65 \times \log^2(Re_{\tau})
\end{equation}

The corresponding normalized mean velocity squared magnitude at the peak is given by:

\begin{equation}
\lim_{Re_{\tau}\to\infty} \frac{U_{\rm g,mean}^2(z/\delta \rvert_{\rm peak})}{U_{\tau}^2} = \left( \frac{2}{\kappa_{\rm g,mean}} \log \left[ \frac{C_{\rm g,fluc}}{2 A_{\rm g,fluc}} \right] \right)^2 = 8.65 \times \log^2(Re_{\tau}),
\end{equation}

\noindent which is the same as for the core value, see Equation (\ref{eq:mean_core}).

The AA value of the mean velocity is:

\begin{equation}
\lim_{Re_{\tau}\to\infty} \biggl \langle \frac{U_{\rm g,mean}^2}{U_{\tau}^2} \biggr \rangle_{\rm AA} = \frac{\log^2(Re_{\tau})}{\kappa_{\rm g,mean}^2} = 8.65 \times \log^2(Re_{\tau}),
\end{equation}

\noindent which is again equal to both the core and peak mean velocity values.

\section{Turbulence Intensity}
\label{sec:ti}

The definition of TI is:

\begin{eqnarray}
I(z) &=& \sqrt{\frac{\overline{u^2_{\rm g,fluc}}(z)}{U_{\rm g,mean}^2(z)}} = \sqrt{\frac{{\overline{u^2_{\rm g,fluc}}}(z)}{U_{\tau}^2} \Biggr / \frac{U_{\rm g,mean}^2(z)}{U_{\tau}^2}} \\
     &=& \sqrt{\frac{B_{\rm g,fluc} - A_{\rm g,fluc} \log (z/\delta) - C_{\rm g,fluc} (z^+)^{-1/2}}{\left( \frac{1}{\kappa_{\rm g,mean}} \log(z^+) + A_{\rm g,mean} \right)^2}},
\end{eqnarray}

\noindent with an asymptotic value of:

\begin{equation}
\lim_{Re_{\tau}\to\infty} I(z) = \frac{\kappa_{\rm g,mean}}{\log(Re_{\tau})} \times \sqrt{\frac{{\overline{u^2_{\rm g,fluc}}}(z)}{U_{\tau}^2}}
\end{equation}

\begin{figure}[!ht]
\includegraphics[width=12cm]{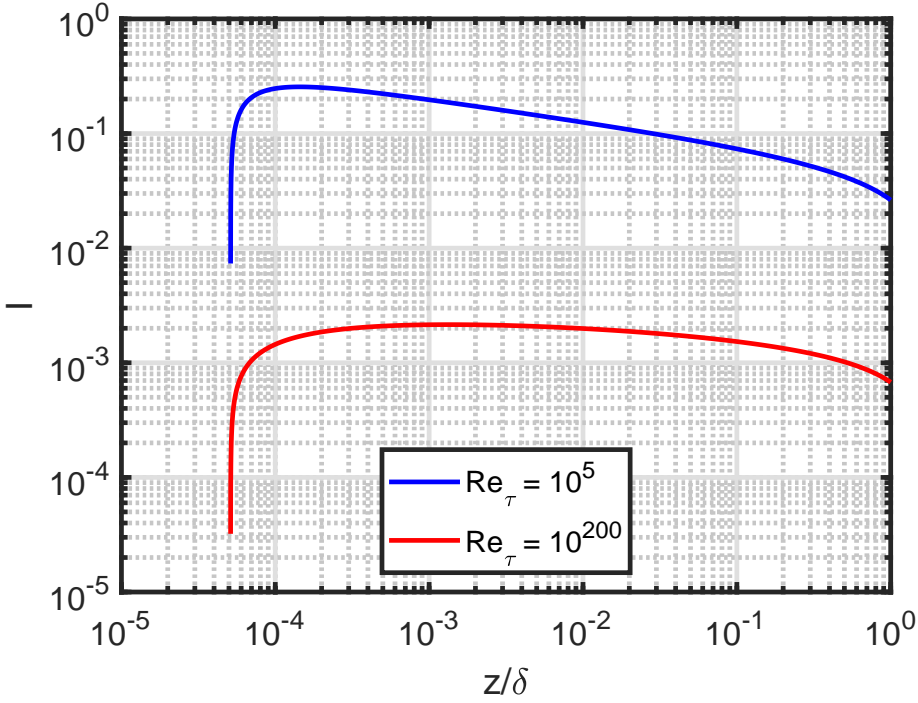}
\caption{TI as a function of $z/\delta$ for the complete expression using asymptotic constants. The blue (red) line is for an $Re_{\tau}$ of $10^5$ ($10^{200}$), respectively.}
\label{fig:I_prof_lo_hi}
\end{figure}

Examples of the TI profile for the minimum and maximum $Re_{\tau}$ treated are shown in Figure \ref{fig:I_prof_lo_hi}.

\subsection{Core}

The asymptotic scaling of the core TI is given by:

\begin{equation}
\lim_{Re_{\tau}\to\infty} I(z/\delta=1) = \frac{\kappa_{\rm g,mean}}{\log(Re_{\tau})} \times \sqrt{0.84} = \frac{0.31}{\log(Re_{\tau})}
\end{equation}

\begin{figure}[!ht]
\includegraphics[width=12cm]{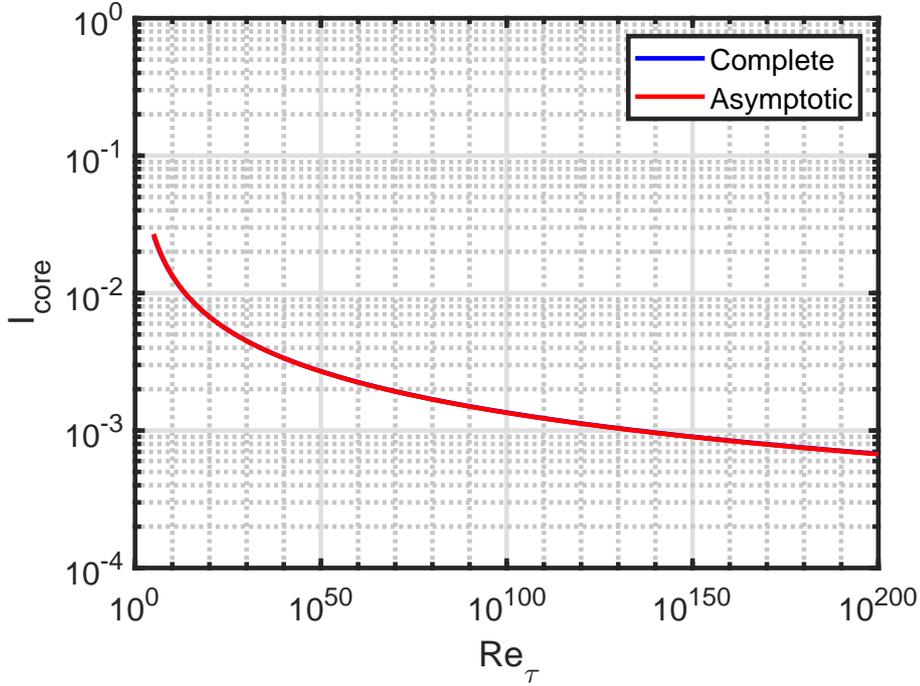}
\caption{Core TI as a function of $Re_{\tau}$. The blue (red) line is for the complete (asymptotic) expression, respectively. The two lines are almost identical, therefore the blue line is not visible.}
\label{fig:I_core}
\end{figure}

The complete and asymptotic core TI scalings are compared in Figure \ref{fig:I_core}. Both scalings are almost indistinguishable from the lowest $Re_{\tau}$ and up.

\subsection{Peak}

\begin{figure}[!ht]
\includegraphics[width=12cm]{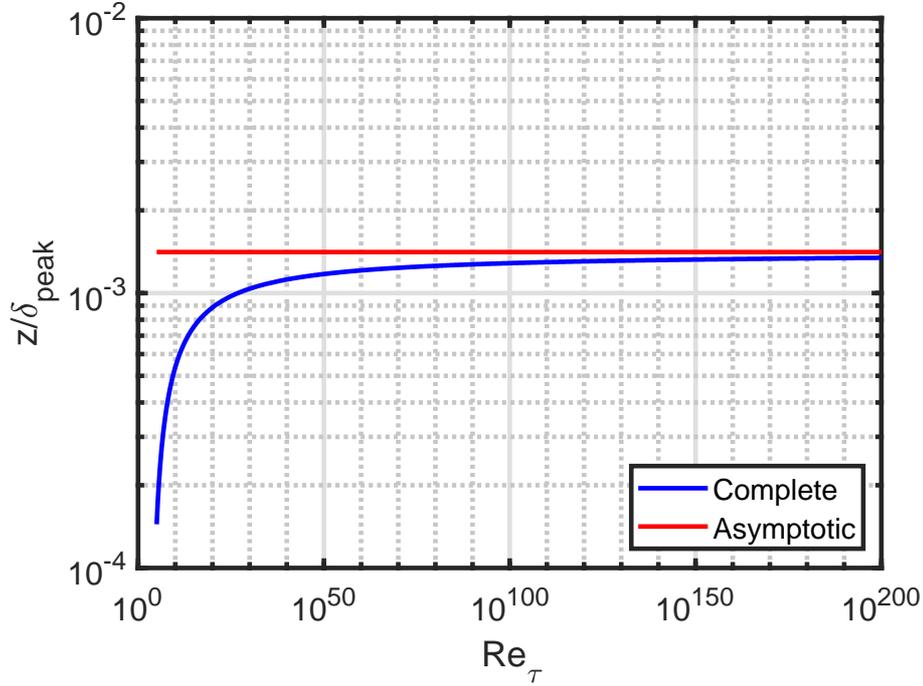}
\caption{$\frac{z}{\delta} \big \rvert_{\rm peak}$ as a function of $Re_{\tau}$. The blue (red) line is for the complete (asymptotic) expression, respectively.}
\label{fig:peak_pos}
\end{figure}

The complete and asymptotic peak TI position are compared in Figure \ref{fig:peak_pos}. We note that the convergence of the complete to the asymptotic scaling is quite slow, differences remain up to the largest $Re_{\tau}$.

The asymptotic scaling of the peak TI is given by:

\begin{equation}
\lim_{Re_{\tau}\to\infty} I(z/\delta \rvert_{\rm peak})) = \frac{\kappa_{\rm g,mean}}{\log(Re_{\tau})} \times \sqrt{8.27} = \frac{0.98}{\log(Re_{\tau})}
\end{equation}

\begin{figure}[!ht]
\includegraphics[width=12cm]{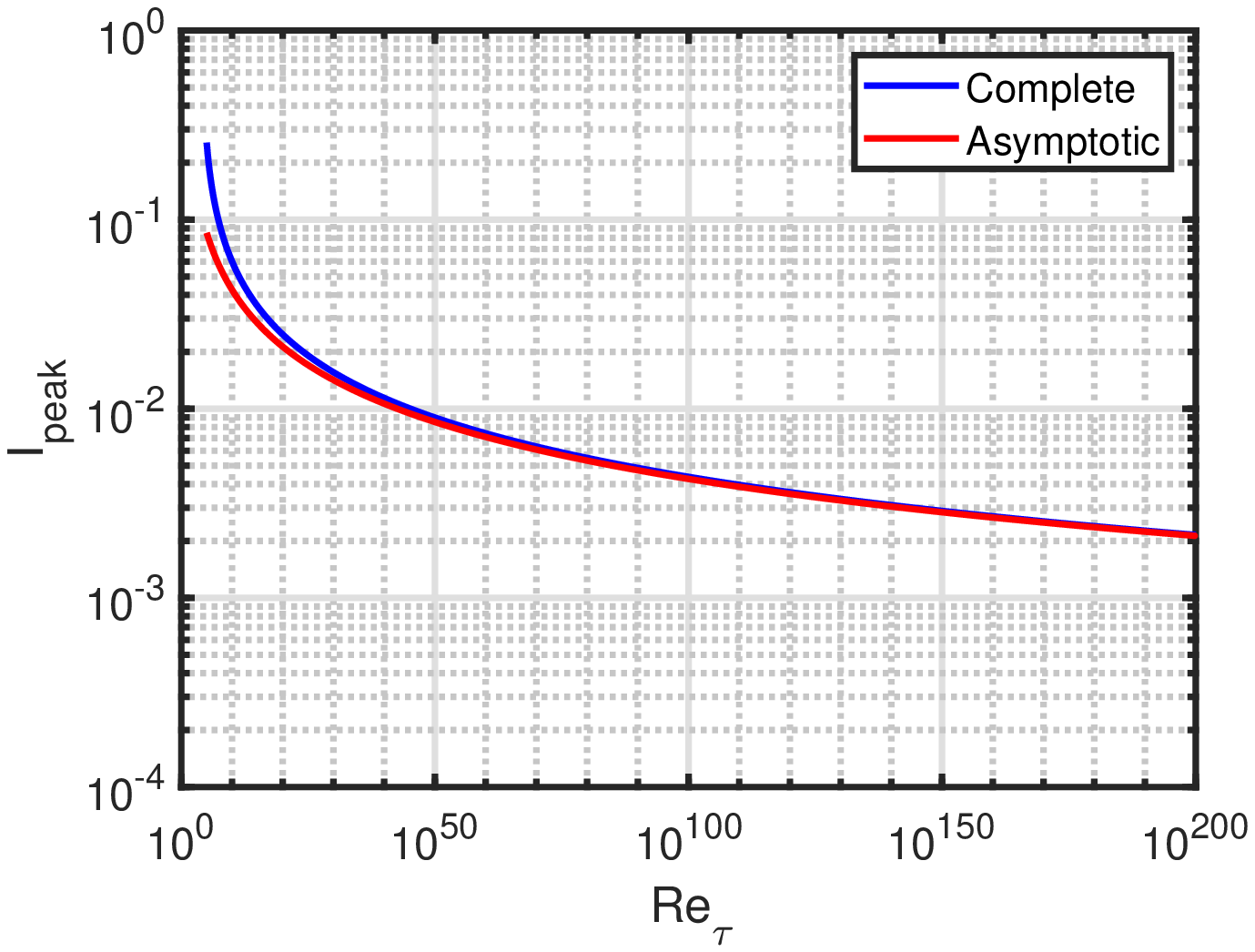}
\caption{Peak TI as a function of $Re_{\tau}$. The blue (red) line is for the complete (asymptotic) expression, respectively.}
\label{fig:I_peak}
\end{figure}

The complete and asymptotic peak TI scalings are compared in Figure \ref{fig:I_peak}. As for the peak position, the complete scaling solution converges to the asymptotic solution relatively slowly, but not as slow as for the peak position.

\subsection{Area-averaged}

The asymptotic scaling of the AA TI is given by:

\begin{equation}
\lim_{Re_{\tau}\to\infty} \langle I(z) \rangle_{\rm AA} = \frac{\kappa_{\rm g,mean}}{\log(Re_{\tau})} \times \sqrt{3.04} = \frac{0.59}{\log(Re_{\tau})}
\end{equation}

\begin{figure}[!ht]
\includegraphics[width=12cm]{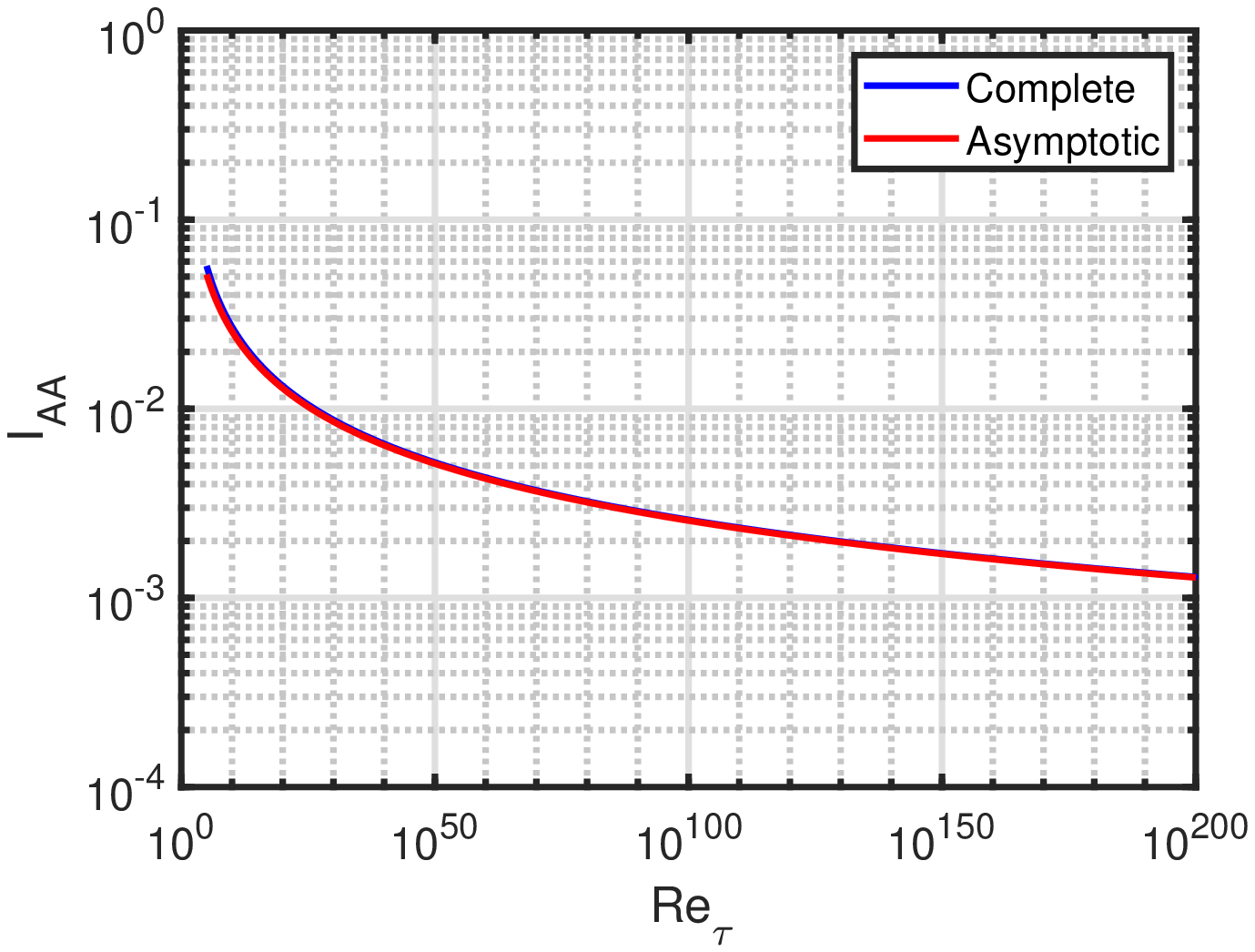}
\caption{AA TI as a function of $Re_{\tau}$. The blue (red) line is for the complete (asymptotic) expression, respectively. The two lines are almost identical, therefore the blue line is only visible for the very lowest $Re_{\tau}$.}
\label{fig:I_AA}
\end{figure}

The complete and asymptotic AA TI scalings are compared in Figure \ref{fig:I_AA}. A slight difference can be seen at the lowest $Re_{\tau}$, but it quickly disappears with increasing $Re_{\tau}$.

\section{Discussion}
\label{sec:disc}

\subsection{Turbulence intensity and the friction factor}

We discuss the relationship between the TI and friction factor and begin by stating the relationship between the friction and bulk Reynolds numbers:

\begin{equation}
Re_{\tau} = \sqrt{\frac{\lambda}{32}} \times Re_D,
\end{equation}

\noindent which can be fit to the Princeton Superpipe measurements \cite{basse_c}:

\begin{eqnarray}
  Re_{\tau} &=& 0.0621 \times Re_D^{0.9148} \\
  Re_D &=& \left( \frac{Re_{\tau}}{0.0621} \right)^{1/0.9148}
\label{eq:re_d_re_tau}
\end{eqnarray}

It is important to emphasize that this relation is derived from measurements with minimum (maximum) $Re_{\tau}$ of 1985 (98190), respectively. It is not clear how well this captures the behavior for higher Reynolds numbers than the measured maximum.

Complete and asymptotic expressions for the smooth pipe friction factor are \cite{basse_c}:

\begin{equation}
\frac{1}{\sqrt{\lambda}} = \frac{1.930 \times \log(Re_D \sqrt{\lambda})}{\log(10)} - 0.537
\end{equation}

\begin{equation}
\lim_{Re_{\tau}\to\infty} \frac{1}{\sqrt{\lambda}} = \frac{1.930}{\log(10) 0.9148} \times \log(Re_{\tau}) = 0.92 \times \log(Re_{\tau})
\end{equation}

The friction factor and TI are related to each other as argued in \cite{tennekes_a}:

\begin{equation}
I \sim \sqrt{\frac{\lambda}{2}} = 0.71 \times \sqrt{\lambda},
\end{equation}

\noindent or:

\begin{equation}
\frac{I^2}{\lambda} \sim \frac{1}{2}
\end{equation}

We note that our AA definition is what comes closest to this estimate:

\begin{equation}
\lim_{Re_{\tau}\to\infty} \frac{\langle I^2(z) \rangle_{\rm AA}}{\lambda} = \frac{3.04 \times \kappa_{\rm g,mean}^2}{(0.92)^2} = 0.42,
\end{equation}

\noindent consistent with the estimates of 0.39 \cite{basse_a} and 0.38 \cite{basse_b}.

\subsection{Comparison to power-law scaling expressions}

We have previously derived both TI core and AA smooth pipe power-law scalings based on the Princeton Superpipe measurements \cite{russo_a,basse_d}:

\begin{equation}
I_{\rm power-law}(z/\delta=1) = 0.0550 \times Re_D^{-0.0407}
\end{equation}

\begin{equation}
\langle I_{\rm power-law} \rangle_{\rm AA} = 0.317 \times Re_D^{-0.110}
\end{equation}

These expressions can be converted to corresponding $Re_{\tau}$-scalings using Equation (\ref{eq:re_d_re_tau}):

\begin{equation}
I_{\rm power-law}(z/\delta=1) = 0.049 \times Re_{\tau}^{-0.044}
\end{equation}

\begin{equation}
\langle I_{\rm power-law} \rangle_{\rm AA} = 0.23 \times Re_{\tau}^{-0.12}
\end{equation}

\begin{figure}[!ht]
\includegraphics[width=6.5cm]{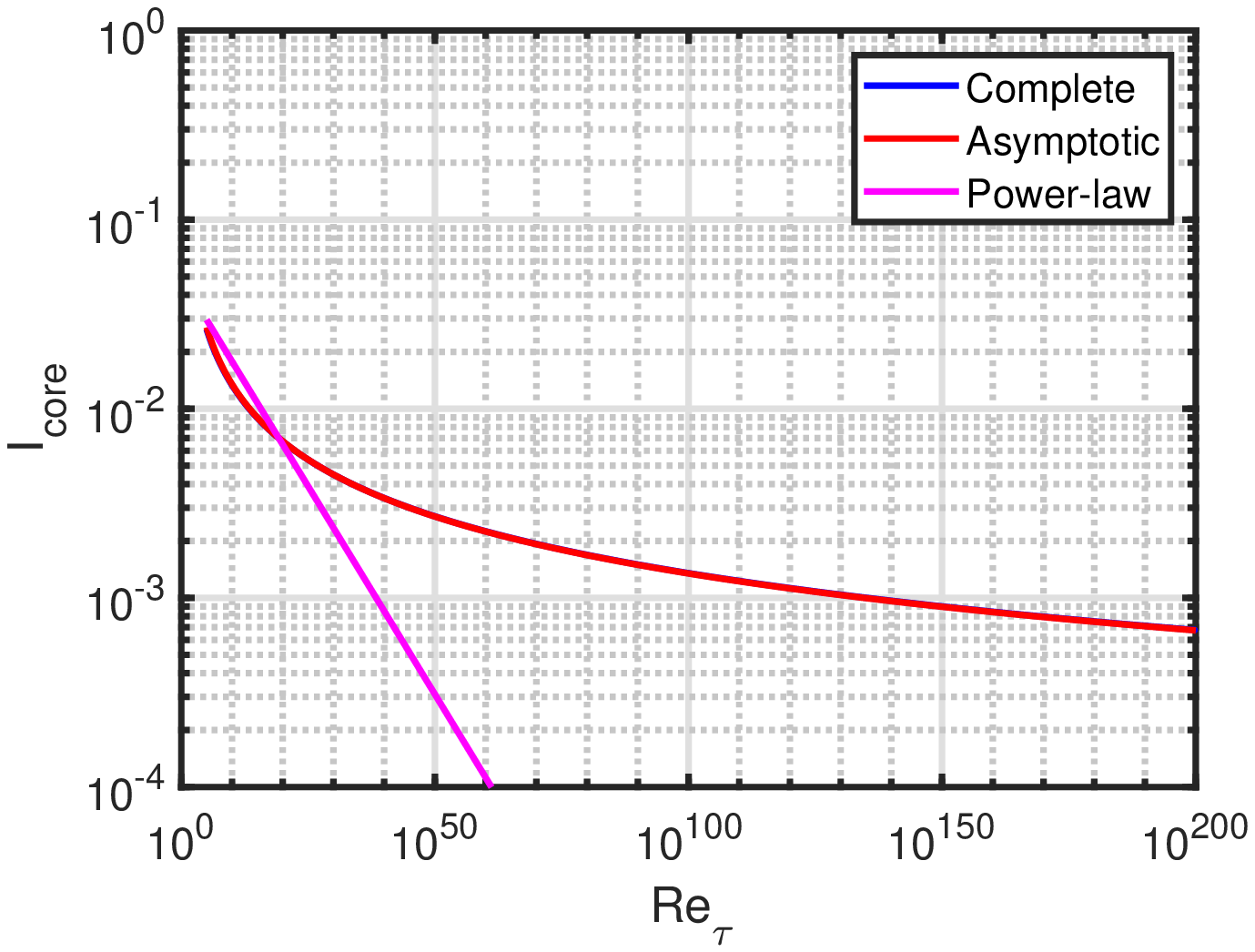}
\includegraphics[width=6.5cm]{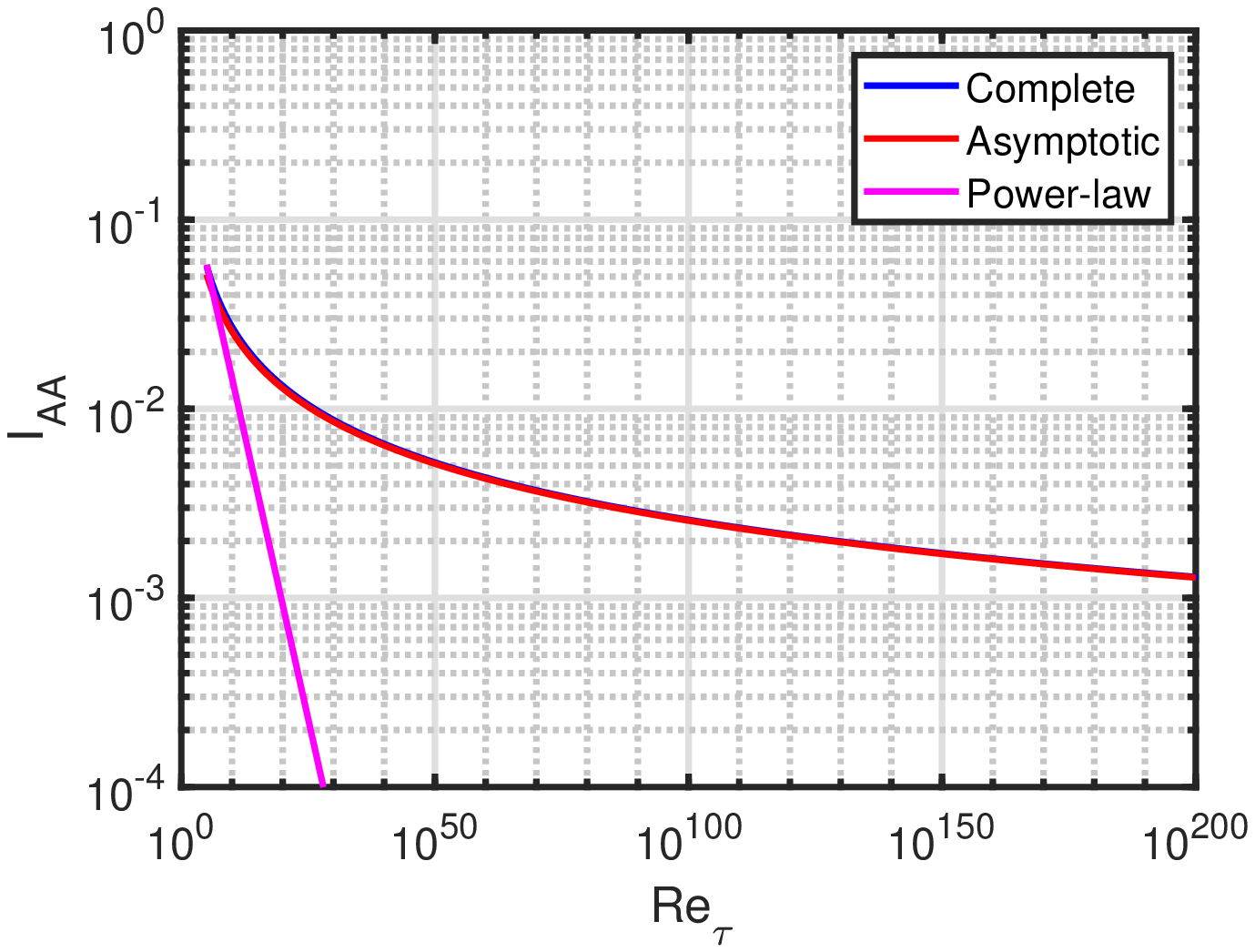}
\caption{TI as a function of $Re_{\tau}$. The blue/red/magenta lines are for the complete/asymptotic/power-law expressions, respectively. Left-hand plot: Core, right-hand plot: AA.}
\label{fig:I_core_AA_plaw}
\end{figure}

The power-law scaling laws are shown in Figure \ref{fig:I_core_AA_plaw} for both the core and AA definitions in addition to the complete and asymptotic scaling expressions shown previously in Figures \ref{fig:I_core} and \ref{fig:I_AA}: It is clear that the power-laws decrease much faster with increasing $Re_{\tau}$.

Since the power-law scalings were derived without considering a transition, we would argue that they are less likely to be correct compared to the ones based on the asymptotic fit parameters. Thus, we recommend that the power-law scalings should only be used for the $Re_{\tau}$-range where they were derived, i.e. from 1985 to 98190.

\section{Conclusions}
\label{sec:conc}

Under the assumption that no further transitions exist beyond $Re_{\tau} \sim 10^5$, we have extrapolated TI scaling to $Re_{\tau} \gg 10^5$. The relationship peak:area-averaged:core TI scaling is 3.16:1.90:1.00 and they all decrease with $1/\log(Re_{\tau})$ due to the mean velocity scaling. The asymptotic scalings for the area-averaged and core scalings can be used for all $Re_{\tau}$ studied, but the peak scaling deviates from the complete scaling for the lower $Re_{\tau}$ range. It is important to note that we have based our scaling expressions on incompressible measurements with a Mach number below 0.2. We are not aware of corresponding research for compressible flow. A comparison between the power-law and log-law scalings shows that they diverge for sufficiently high Reynolds numbers.

Asymptotically, we confirm the scaling $I^2/\lambda \sim 1/2$ for the AA definition, but with a constant of around 0.4 instead of 0.5.

\paragraph{Acknowledgements}

We thank Professor Alexander J. Smits for making the Princeton Superpipe data publicly available.

\paragraph{Data availability statement}

Data sharing is not applicable to this article as no new data were created or analyzed in this study.

\clearpage

\label{sec:refs}

\end{document}